\documentclass[11pt,twoside]{article}
\usepackage{asp2010}

\resetcounters

\allowtitlefootnote

\begin{document}

\title{Imaging Planetary Nebulae with Herschel-PACS and SPIRE $^\ast$}

\author{P.A.M.~van~Hoof$^{1}$,
  K.M.~Exter$^{3}$,
  G.C.~Van~de~Steene$^{1}$,
  M.J.~Barlow$^{2}$,
  T.L.~Lim$^{8}$,
  B.~Sibthorpe$^{4}$,
  M.A.T.~Groenewegen$^{1}$,
  T.~Ueta$^{5}$,
  M.~Matsuura$^{2}$,
  J.A.D.L.~Blommaert$^{3}$,
  M.~Cohen$^{6}$,
  W.~De~Meester$^{3}$,
  W.K.~Gear$^{7}$,
  H.L.~Gomez$^{7}$,
  P.C.~Hargrave$^{7}$,
  E.~Huygen$^{3}$,
  R.J.~Ivison$^{4}$,
  C.~Jean$^{3}$,
  S.J.~Leeks$^{8}$,
  G.~Olofsson$^{9}$,
  E.T.~Polehampton$^{8,10}$,
  S.~Regibo$^{3}$,
  P.~Royer$^{3}$,
  B.M.~Swinyard$^{8}$,
  B.~Vandenbussche$^{3}$,
  H.~Van~Winckel$^{3}$,
  C.~Waelkens$^{3}$,
  R.~Wesson$^{2}$
\affil{$^{1}$Royal Observatory of Belgium, Ringlaan 3, B-1180 Brussels, Bel\-gium}
\affil{$^{2}$Dept.\ of Physics \& Astronomy, UCL, Gower St, London WC1E 6BT, UK}
\affil{$^{3}$IvS, Katholieke Univ.\ Leuven, Ce\-les\-tij\-nenlaan 200 D, Leuven, Belgium}
\affil{$^{4}$UK Astronomy Technology Centre, ROE, Blackford Hill, Edinburgh, UK}
\affil{$^{5}$Dept. of Physics and Astronomy, Univ.\ of Denver, Denver, CO 80208, USA}
\affil{$^{6}$Radio Astronomy Laboratory, Univ.\ of California at Berkeley, CA 94720, USA}
\affil{$^{7}$School of Physics and Astronomy, Cardiff University, Cardiff, Wales, UK}
\affil{$^{8}$Rutherford Appleton Laboratory, Oxfordshire, OX11 0QX, UK}
\affil{$^{9}$Dept.\ of Astronomy, Stockholm University, Stockholm, Sweden}
\affil{$^{10}$Department of Physics, Univ.\ of Lethbridge, Alberta, T1J 1B1, Canada}
}

\begin{abstract}
  In this paper we will discuss the images of Planetary Nebulae that have
  recently been obtained with PACS and SPIRE on board the Herschel satellite.
  This comprises results for NGC 650 (the little Dumbbell nebula), NGC 6853
  (the Dumbbell nebula), and NGC 7293 (the Helix nebula).
\end{abstract}

\footnotetext{$^\ast$ Herschel is an ESA space observatory with science instruments
  provided by European-led Principal Investigator consortia and with important
  participation from NASA.}

\section{Introduction}

We have obtained Herschel PACS and SPIRE images of planetary nebulae (PNe) as
part of the MESS (Mass loss of Evolved StarS) guaranteed time key program. The
aims of this program are threefold: 1) study the time dependence of the mass
loss process via a search for shells and multiple shells, 2) study the dust
and gas chemistry as a function of progenitor mass, and 3) study the
properties and asymmetries of a representative sample of evolved objects. This
program will be discussed in more detail in a forthcoming paper \citep{Gro10}.
During the science demonstration phase of Herschel we obtained PACS and SPIRE
images of NGC 6720, which have been discussed in \citet{vH10}. The detailed
match between the H$_2$ and dust emission in this object appears to be the
first observational evidence that H$_2$ forms on oxygen-rich dust. The most
plausible scenario is that the H$_2$ resides in high density knots that were
formed after the recombination of the gas started when the central star
entered the cooling track. Models indicate that a substantial amount of H$_2$
could have formed since that time and that the formation may still be ongoing
at this moment.

Below we will present preliminary reductions of PACS and SPIRE data of 3 other
PNe that we obtained during the routine phase.

\section{Data Reduction}

\begin{figure}
\mbox{\centerline{\includegraphics[width=0.8\textwidth]{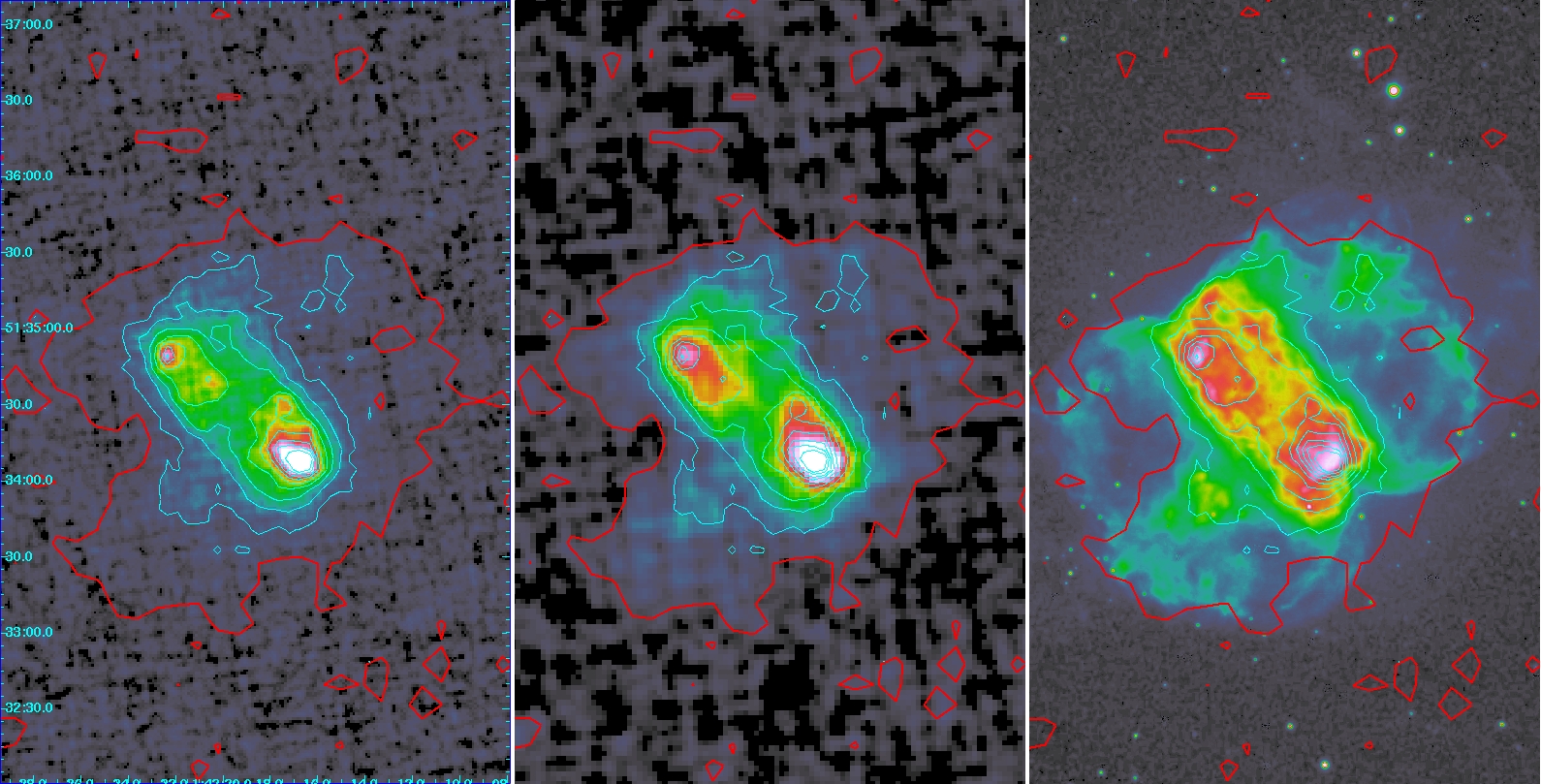}}}
\caption{NGC 650. Left: PACS 70~$\mu$m, middle PACS 160~$\mu$m, right:
  ground-based H$\alpha$ image (Lars \O. Andersen, Lars Malmgren, Frank R.
  Larsen, NOT). A manual shift of 4.5\arcsec\ was applied to the WCS of the
  PACS images.\label{ngc650}}
\end{figure}

\begin{figure}
\mbox{\centerline{\includegraphics[width=0.8\textwidth]{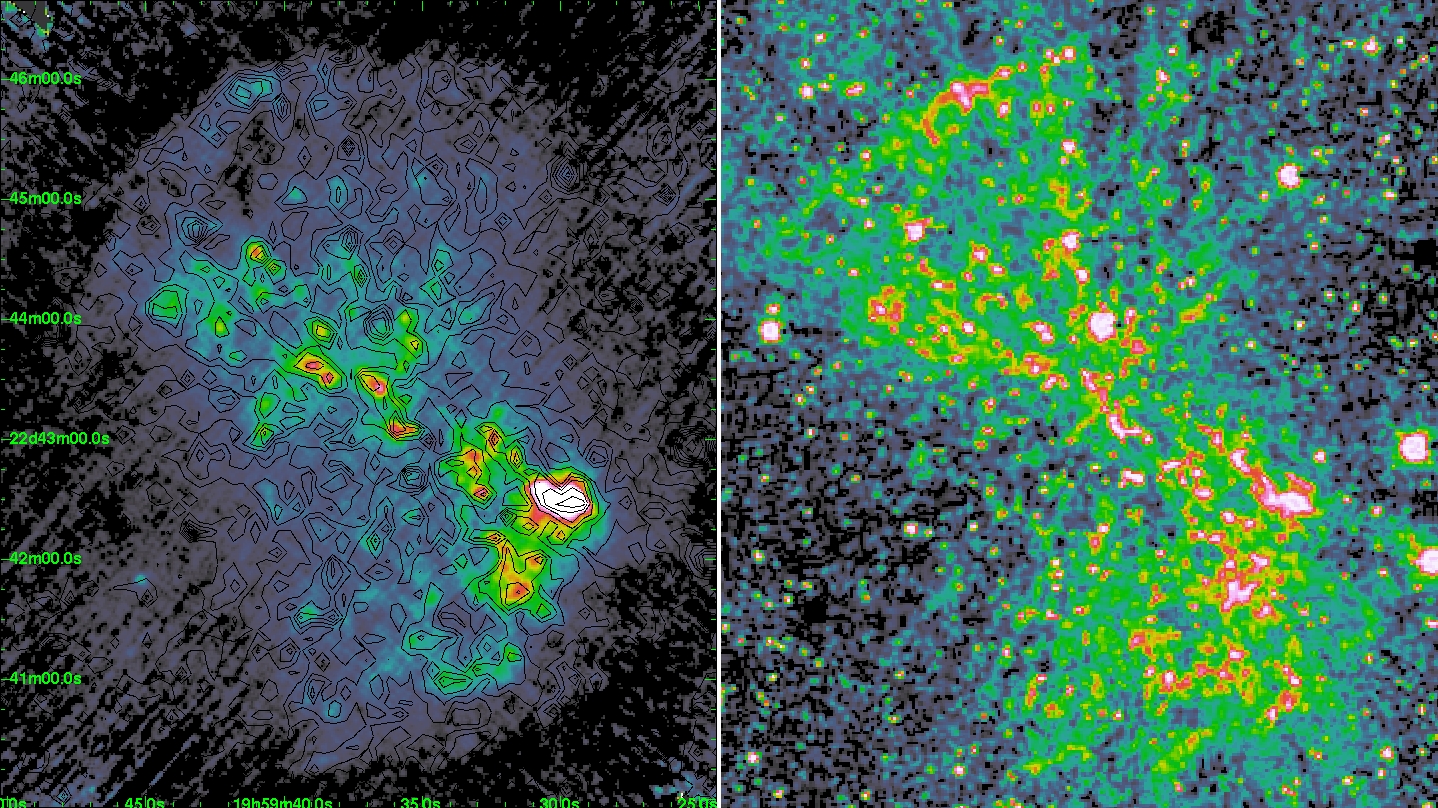}}}
\caption{NGC 6853. Left: PACS 70~$\mu$m with H$_2$ contours overlaid, right:
  ground-based H$_2$ 2.12~$\mu$m image (I. Gatley, M. Merrill,
  NOAO).\label{ngc6853}}
\end{figure}

\citet{Po10} describe the standard data reduction scheme for PACS scan maps
from so-called ``Level 0'' (raw data) to ``Level 2'' products. The level 1 and
2 products that are part of the data sets from the Herschel science archive
have been produced by execution of these standard pipeline steps. We do not
use the standard products, but use an adapted and extended version of the
pipeline script suited to our needs starting from the raw data. In particular
the deglitching step(s), and the ``high pass filtering'' need special care.

When the central source is bright, the deglitching task incorrectly masks a
significant number of frames of the source as glitches. A second pass (called
``2nd level deglitching'' is needed to remedy this. The purpose of the high
pass filter is to remove the $1/f$ noise from the images. At the moment the
task is using a median filter, which subtracts a running median from each
readout. This works well for point sources, but causes significant problems
for extended sources (resulting in negative ``shadows'' in the image). To
prevent the artifacts, any part of the source needs to be masked from the
median filter. We are currently still experimenting with various algorithms to
achieve this in an optimal way. We are also looking into using MadMap as an
alternative algorithm. In addition, as the standard pipeline script operates
on a single Astronomical Observation Request (AOR), while our observations are
always the concatenation of 2 AORs (a scan and a cross-scan) an additional
step is also needed to combine the two scans. The pointing accuracy of
Herschel is within specifications, but is however not negligible compared to
the PACS beam size. Hence manual adjustments of the world coordinate system
(WCS) may also be necessary when comparing with other (groundbased) images. A
more in-depth discussion of the PACS data reduction will be given in
\citet{Gro10}.

\begin{figure}
\mbox{\centerline{\includegraphics[width=0.9\textwidth]{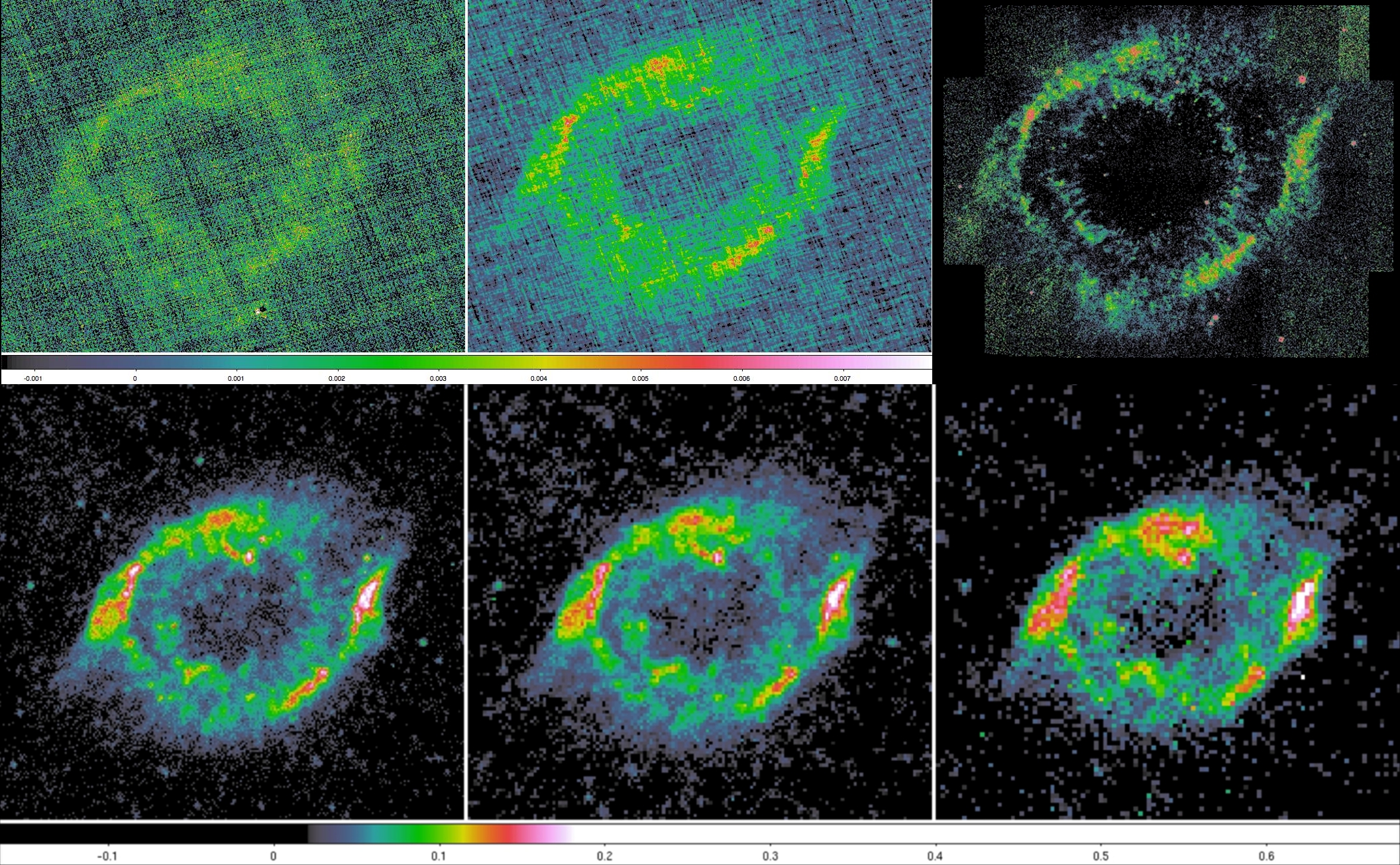}}}
\caption{NGC 7293. Top row: PACS 70 \& 160~$\mu$m, H$_2$ 2.12~$\mu$m (courtesy
  A. Speck, KPNO), bottom row SPIRE 250, 350 \& 500~$\mu$m.\label{ngc7293}}
\end{figure}

The standard SPIRE photometer data processing pipeline, described by
\citet{Gr08, Gr10}, is sufficient to reduce our SPIRE photometric imaging
data. The calibration steps for these data are described by \citet{Sw10}.

In Figs.~\ref{ngc650}, \ref{ngc6853}, and \ref{ngc7293} we present preliminary
reductions of the PACS scan maps of NGC650, NGC 6853, and NGC 7293. The
reduction of the SPIRE maps of NGC 7293 is final.

\section{Outlook}

Both NGC 7293 and NGC 6853 are well evolved PNe with central stars that are
currently on the cooling track \citep{OD07}. So both nebulae are in a similar
evolutionary state compared to NGC 6720. Especially NGC 7293 is almost an
``older twin''. Also in these objects we see that the distribution of the
H$_2$ closely follows the dust. This suggests that in these objects the H$_2$
has also been formed in high density knots. We will model these PNe in detail
to test whether the scenario that we proposed in \citet{vH10} can also explain
the H$_2$ formation in these objects. Also in NGC 650 we see that the gas and
dust show a very similar distribution. It is not yet clear however if this
H$_2$ formed in a similar fashion, or whether it is leftover AGB material.

In addition to the scan maps presented in \citet{vH10} and this paper, we will
also obtain Herschel images of NGC 650 (SPIRE), NGC 3587 (PACS, SPIRE), NGC
6543 (PACS, SPIRE), NGC 6853 (SPIRE), NGC 7027 (PACS), IRAS 22036$+$5306
(PACS). For all targets we will obtain PACS images in the 70 and 160~$\mu$m
bands, while SPIRE images will be obtained in all 3 bands: 250, 350, and
500~$\mu$m. The final goal of the program is to study the structures of their
dust shells in order to learn about the (post-) AGB mass loss processes.

\acknowledgements

PvH, KE, GVdS, MG, JB, WDM, RH, CJ, SR, PR, and BV acknowledge support from
the Belgian Federal Science Policy Office via the PRODEX Programme of ESA.
PACS has been developed by a consortium of institutes led by MPE (Germany) and
including UVIE (Austria); KU Leuven, CSL, IMEC (Belgium); CEA, LAM (France);
MPIA (Germany); INAFIFSI/ OAA/OAP/OAT, LENS, SISSA (Italy); IAC (Spain). This
development has been supported by the funding agencies BMVIT (Austria),
ESA-PRODEX (Belgium), CEA/CNES (France), DLR (Germany), ASI/INAF (Italy), and
CICYT/MCYT (Spain). SPIRE has been developed by a consortium of institutes led
by Cardiff Univ. (UK) and including Univ. Lethbridge (Canada); NAOC (China);
CEA, LAM (France); IFSI, Univ. Padua (Italy); IAC (Spain); Stockholm
Observatory (Sweden); Imperial College London, RAL, UCL-MSSL, UKATC, Univ.
Sussex (UK); Caltech, JPL, NHSC, Univ. Colorado (USA). This development has
been supported by national funding agencies: CSA (Canada); NAOC (China); CEA,
CNES, CNRS (France); ASI (Italy); MCINN (Spain); SNSB (Sweden); STFC (UK); and
NASA (USA). Data presented in this paper were analysed using ``HIPE'', a joint
development by the Herschel Science Ground Segment Consortium, consisting of
ESA, the NASA Herschel Science Center, and the HIFI, PACS and SPIRE consortia.
This research made use of tools provided by Astrometry.net.

\bibliography{vanHoof}

\begin{thebibliography}{}
\expandafter\ifx\csname natexlab\endcsname\relax\def\natexlab#1{#1}\fi
\expandafter\ifx\csname url\endcsname\relax
  \def\url#1{\texttt{#1}}\fi
\expandafter\ifx\csname urlprefix\endcsname\relax\def\urlprefix{URL }\fi
\providecommand{\eprint}[2][]{\url{#2}}

\bibitem[{{Griffin} et~al.(2008){Griffin}, {Dowell}, {Lim}, \& {others}}]{Gr08}
{Griffin}, M., {Dowell}, C.~D., {Lim}, T., \& {others} 2008, Proc.\ of the
  SPIE, Vol.\ 7010, 70102Q

\bibitem[{{Griffin} et~al.(2010){Griffin}, {Abergel}, {Ade}, \&
  {others}}]{Gr10}
{Griffin}, M.~J., {Abergel}, A., {Ade}, P.~A.~R., \& {others} 2010, \aap, 518,
  L3

\bibitem[{{Groenewegen} et~al.(2010){Groenewegen}, {Waelkens}, {Barlow}, \&
  {others}}]{Gro10}
{Groenewegen}, M.~A.~T., {Waelkens}, C., {Barlow}, M.~J., \& {others} 2010,
  \aap, submitted

\bibitem[{{O'Dell} et~al.(2007){O'Dell}, {Sabbadin}, \& {Henney}}]{OD07}
{O'Dell}, C.~R., {Sabbadin}, F., \& {Henney}, W.~J. 2007, \aj, 134, 1679

\bibitem[{{Poglitsch} et~al.(2010){Poglitsch}, {Waelkens}, {Geis}, \&
  {others}}]{Po10}
{Poglitsch}, A., {Waelkens}, C., {Geis}, N., \& {others} 2010, \aap, 518, L2

\bibitem[{{Swinyard} et~al.(2010){Swinyard}, {Ade}, {Baluteau}, \&
  {others}}]{Sw10}
{Swinyard}, B.~M., {Ade}, P.~A.~R., {Baluteau}, J.-P., \& {others} 2010, \aap,
  518, L4

\bibitem[{{van Hoof} et~al.(2010){van Hoof}, {Van de Steene}, {Barlow}, \&
  {others}}]{vH10}
{van Hoof}, P.~A.~M., {Van de Steene}, G.~C., {Barlow}, M.~J., \& {others}
  2010, \aap, 518, L137

\end{thebibliography}

\end{document}